# On the Role of Elastic Strain on Electrocatalysis of Oxygen Reduction Reaction on Pt

Vijay A. Sethuraman,[1,4,*] Deepa Vairavapandian,[1] Manon C. Lafouresse,[2,5] Tuhina A. Maark,[1,6] Naba Karan,[1] Shouheng Sun,[3] Ugo Bertocci,[2] Andrew A. Peterson,[1] Gery R. Stafford,[2,*] Pradeep R. Guduru[1,*]

[1]School of Engineering, Brown University, Providence, Rhode Island 02912, USA
[2]National Institute of Standards and Technology, Materials Measurement Laboratory
100 Bureau Drive, Gaithersburg, Maryland 20899, USA
[3]Department of Chemistry, Brown University, Providence, Rhode Island 02912, USA

## Abstract

The effect of elastic strain on catalytic activity of platinum (Pt) towards oxygen reduction reaction (ORR) is investigated through de-alloyed Pt-Cu thin films; stress evolution in the de-alloyed layer and the mass of the Cu removed are measured in real-time during electrochemical de-alloying of (111)-textured thin-film PtCu (1:1, atom %) electrodes. *In situ* stress measurements are made using the cantilever-deflection method and nano-gravimetric measurements are made using an electrochemical quartz crystal nanobalance. Upon de-alloying *via* successive voltammetric sweeps between -0.05 and 1.15 V *vs.* standard hydrogen electrode, compressive stress develops in the de-alloyed Pt layer at the surface of thin-film PtCu electrodes. The de-alloyed films also exhibit enhanced catalytic activity towards ORR compared to polycrystalline Pt. *In situ* nanogravimetric measurements reveal that the mass of de-alloyed Cu is approximately 210 ± 46 ng/cm$^2$, which corresponds to a de-alloyed layer thickness of 1.2 ± 0.3 monolayers or 0.16 ± 0.04 nm. The average biaxial stress in the de-alloyed layer is estimated to be 4.95 ± 1.3 GPa, which corresponds to an elastic strain of 1.47 ± 0.4%. In addition, density functional theory calculations have been carried out on biaxially strained Pt (111) surface to characterize the effect of strain on its ORR activity; the predicted shift in the limiting potentials due to elastic strain is found to be in good agreement with the experimental shift in the cyclic voltammograms for the dealloyed PtCu thin film electrodes.

[4]Present address: Department of Materials Engineering, Indian Institute of Science, CV Raman Avenue, Bangalore 560012, India.
[5]Present address: Laboratoire de Photophysique et Photochimie Supramoléculaires et Macromoléculaires, École Normale Supérieure de Cachan, 61 Avenue du President Wilson, 94235 Cachan cedex, France.
[6]Present address: Department of Science and Technology Unit of Nanoscience and Thematic Unit of Excellence, Department of Chemistry, Indian Institute of Technology Madras, Chennai 600036, India.

*Corresponding authors:
Email: Pradeep_Guduru@Brown.edu (Telephone: +1 401 863 3362);
vijay@materials.iisc.ernet.in;
Gery.Stafford@nist.gov

## 1. Introduction

Enhanced catalytic activity of de-alloyed PtCu electrodes towards oxygen reduction (ORR) reaction has been demonstrated in a number of recent investigations in thin-film and core-shell geometries [1-19]. The enhancement in catalytic activity is typically attributed to the electronic interaction between the core and shell constituents as well as to the epitaxial mismatch strain in the metallic shell. For example, Strasser *et al.* [4] demonstrated enhanced catalytic activity of de-alloyed PtCu core-shell nanoparticles towards ORR, which was attributed to the pseudomorphic compressive strain in the Pt-enriched shell. The magnitude of the strain in the Pt-enriched shell was estimated using lattice-constant measurements (*via* X-ray diffraction). In this work, real-time stress and nano-gravimetric measurements during electrochemical de-alloying of thin-film PtCu (1:1, atom %) electrodes are reported. *In situ* stress measurements are made using the cantilever-deflection method, and nano-gravimetric measurements are made using an electrochemical quartz crystal nanobalance (EQCN). Upon de-alloying *via* successive voltammetric sweeps between -0.05 and 1.15 V *vs.* standard hydrogen electrode (SHE), compressive stress develops in the de-alloyed layer region near the surface of thin-film PtCu electrodes. *In situ* nanogravimetric measurements reveal that the mass of de-alloyed Cu is approximately 210 ± 46 ng/cm$^2$, which corresponds to a de-alloyed layer thickness of 1.2 ± 0.3 monolayers or 0.16 ± 0.04 nm. From the real-time curvature measurements of the substrate, the average biaxial stress in the de-alloyed layer is estimated to be 4.95 ± 1.3 GPa, which corresponds to an elastic strain of 1.47 ± 0.4%. We have also carried out density functional theory (DFT) calculations on biaxially strained Pt (111) surface to characterize the effect of strain on its ORR activity; the results are found to be in reasonably good agreement with the experimental measurement. Section 2 describes the experimental methods used for real-time stress measurements and EQCN measurement of de-alloyed Cu. Section 3 describes the computational approach to estimate the effect of elastic strain on the expected change in catalytic activity, followed by Section 4 that presents the results and discussion.

## 2. Experimental methods

*Electrode preparation and characterization* – Pt thin films are fabricated by direct-current sputtering (Lesker Lab 18 modular thin-film deposition system, Jefferson Hills, PA) of a platinum target (2” diameter, 0.125” thick disc, 99.995% Pt, Kurt J. Lesker Company) at 100 W power and a pressure of 0.27 Pa of Argon. PtCu thin films are fabricated by co-sputtering Pt and Cu targets (2” diameter, 0.25” thick disc, 99.995% Cu, Kurt J. Lesker Company) each at 100 W power and at a pressure of 0.27 Pa of Argon (99.995%). Film thicknesses are monitored by a quartz crystal microbalance during the sputter deposition process and independently verified by white-light interferometry. The composition (*i.e.,* atomic ratio) of the PtCu films is measured by Auger electron spectroscopy (AES) as well as energy dispersive X-ray spectroscopy (EDS). For all of our studies on the PtCu alloy, the atomic ratio of Pt and Cu is approximately 1:1 (± 2%). The following three different types of substrates are used: (1) Ti/Au-coated quartz crystals (1” diameter, 5 MHz resonance frequency, Inficon *Inc.*, Newton, MA), (2) Ti/Au-coated borosilicate glass cantilevers (100 μm thick, 3 mm wide, and 60 mm long, Precision Glass and Optics, Santa Ana, CA), and (3) Polished glassy carbon (GC) cylindrical discs (5 mm diameter, 4 mm thick , Pine Research Instrumentation, Grove City, PA). The PtCu films on quartz crystals are used to accurately measure the mass of Cu removed during the de-alloying process; the Pt and PtCu films on borosilicate cantilevers are used to make direct measurements of stress evolution during the de-alloying process while also measuring changes in catalytic activity; and the films on GC substrates are used in a rotating-disc-electrode (RDE) set up to study ORR under well-defined hydrodynamic conditions. The experimental parameters are: rotation speed of 1200 rpm, a temperature of 298 K; and at a scan rate of 20 mV/s. The crystallinity and the surface composition of the sputter-deposited PtCu films are characterized by X-ray diffraction (XRD) and X-ray photoelectron spectroscopic (XPS) measurements, respectively. The surface roughness of the PtCu film is measured by an atomic-force microscope (AFM).

*Electrochemical de-alloying of PtCu* – As demonstrated in the literature [1-6], de-alloying is carried out by electrochemical means – *via* cyclically scanning the potential of the PtCu electrode between -0.2 and +0.8 V *vs.* SHE at a scan rate of 20 mV/s (for GC electrodes). In all cases, 0.1 M perchloric acid ($HClO_4$) saturated with either oxygen or argon (both UHP grade) is used as the electrolyte.  A standard three-electrode electrochemical cell with an Ag/AgCl reference electrode and a platinum-mesh counter electrode is used.

*In situ stress measurements via cantilever-deflection* – The stress evolution in the Pt-enriched layer, which forms during the de-alloying of PtCu, is measured *in situ* by continuously tracking the curvature of the borosilicate-glass substrate with reference to its initial value [20-32]. A schematic of the experimental setup is shown in Figure 1(a), which shows the electrochemical cell and the setup for the cantilever-deflection measurement. A HeNe laser beam (632.8 nm, 0.5 mW, 0.48 mm beam diameter, JDS Uniphase Model 1108P) is reflected from the cantilever surface and its position is monitored using a duo-lateral position-sensitive detector (PSD, 20 mm x 20 mm active area, DLS-20, UTD Sensors) to measure the cantilever deflection.  A small Pyrex cell (15 ml volume) with Pt-plate counter electrode and saturated sulfate reference electrode (SSE) is used to conduct the electrochemical experiments. The electrolyte (0.1 M $HClO_4$) is purged with UHP argon prior to the de-alloying experiment; argon purge above the electrolyte is continued during the de-alloying experiment. An EG&G 273 potentiostat (Princeton Applied Research, Oak Ridge, TN) is used to electrochemically de-alloy Cu from PtCu electrodes. Ti/Au-coated (10/200 nm) borosilicate glass cantilevers serve as substrates for the 100 nm PtCu thin-film electrodes (Figure 1b).  The purpose of the Au layer is to serve as an electrochemically stable current collector, and to obtain (111)-textured PtCu films; the deposition conditions are controlled to yield predominantly (111)-oriented Au films, which can be seen in the XRD spectrum in Figure 2.

The product of the stress in the de-alloyed Pt layer, $\sigma_{Pt}$, and its thickness, $\mathrm{h}_{\mathrm{Pt}}$, on the sample (commonly referred to as *stress-thickness*) is related to the curvature of the substrate through

$$\sigma_{Pt}\mathrm{h}_{\mathrm{Pt}} = \frac{E_s h_s^2}{6(1-\upsilon_s)R} \quad where \quad \frac{1}{R} = \frac{n_{air} d_{psd}}{2 L n_{H_2O} D_{psd}} \qquad 1$$

where $E_s$, $\upsilon_s$, and $h_s$ are the Young's modulus, Poisson ratio, and thickness of the borosilicate-glass substrate, respectively, and $R$ is the radius of curvature of the cantilever. Changes in the cantilever curvature were monitored by tracking the location of the reflected laser beam along the vertical axis of the PSD. A small angle approximation was used to evaluate the curvature of the glass cantilever directly from the laser spot displacement on the PSD, see Eqn. 1, where $\boldsymbol{d_{psd}}$ is the change in the vertical coordinate of the reflected laser beam onto the PSD, $\boldsymbol{D_{psd}}$ is the distance of the PSD from the electrode, and $\boldsymbol{L}$ is the length of the cantilever, measured from the point where the cantilever is clamped to where the laser strikes the surface; $n_{air}$ and $n_{H_2O}$ are the refractive indices of air and the electrolyte, respectively [20, 29, 33, 34]. Values of the parameters in the above equation are given in Table 1. To obtain the value of the stress ($\sigma_{Pt}$) in the de-alloyed layer, an independent calculation of the thickness ($\mathrm{h_{Pt}}$) of the stressed layer is necessary. An electrochemical quartz crystal nanobalance (EQCN, Maxtek *Inc.*) is used to measure the mass of Cu removed during de-alloying, which is used to calculate the thickness of the Pt-enriched layer [26-29, 35].

## 3. Computational methods

In this work, a DFT-based program DACAPO is used with the atomic simulation environment (ASE) to carry out electronic structure calculations [36, 37]. The computational software utilizes plane waves, Vanderbilt ultrasoft pseudopotentials, and the revised Perdew-Burke-Ernzerhof (RPBE) density functional [38-40]. A 2×2 unit cell with four layers and 20 Å vacuum using a RPBE bulk lattice constant (4.02 Å) was used to model the Pt (111) surface. Out of the four layers, the bottom two layers are fixed while the top two layers and the adsorbates are relaxed. All geometry optimizations are carried out using a plane wave cut-off of 450 eV, density cut-off of 500 eV, and 4×4×1 k-point mesh. In the simulation cell $[\bar{1}10]$, $[11\bar{2}]$, and [111] directions form the x, y, and z-axes, respectively. A uniform in-plane biaxial strain ($e_x = e_y$) of up to ±2.5% is applied to the Pt (111) surface. The corresponding induced relaxation along the z-axis was calculated based on elastic stiffness constants ($s_{11}$, $s_{12}$, and $s_{44}$) and was explicitly included while constructing the slab models.

ORR is studied on biaxially strained Pt (111) surfaces *via* both dissociative and associative mechanisms as outlined in Figure 3 [40-41]. Based on the scheme shown in Figure 3, O*, HO*, HOO*, and $O_2$* intermediates are studied, each of which is adsorbed at its most preferred site on unstrained Pt (111) as shown in Figure 4. Note that * represents an adsorption site on the Pt (111) surface. To evaluate free energies at 298 K, contributions for zero-point energy, entropy, and solvent are utilized as described in Refs. [37-40, 42]. For O*, HO* and HOO*, solvent stabilizations of 0.0, -0.60, and -0.23 eV, respectively are considered. To accurately calculate the adsorption free energy of $O_2$* formation, the bond energy of the reference $O_2$ molecule is derived from the RPBE free energies of $H_2O$ and $H_2$ molecules from this work and the experimental reaction free energy (2.46 eV) for $H_2O \rightarrow ½\ O_2 + H_2$. Steps D2, D3, A2, A4, and A5 in the scheme shown in Figure 3 are associated with an $(H^+ + e^-)$ transfer. For a reaction of the form $A^* + (H^+ + e^-) \rightarrow {}^*AH$, the computational hydrogen electrode model [40] is adopted to obtain the reaction free energy of the step as that for $A^* + ½\ H_2 \rightarrow {}^*AH$. A term $\Delta G_U = -e\ U$ is added to adjust the reaction potential to 0 V.

## 4. Results and discussion

A schematic of the PtCu thin-film electrode before and after de-alloying is shown in Figure 5a. The first four CV cycles corresponding to de-alloying of the film are shown in Figure 5b. The Cu dissolution peaks are seen at 0.7 V *vs.* SHE, which are consistent with that reported in the literature [3, 8, 9, 11, 43-46]. Cu dissolution charge (*i.e.*, area under the anodic peak) decreases with cycle number and approaches zero (Figure 5c). The surface composition of as-prepared and de-alloyed PtCu thin-film electrodes are characterized by XPS. Surface spectra (Figure 6) of as-prepared PtCu film show peaks which correspond to $Cu^{2p}$ and $Pt^{4f}$ levels indicating the presence of Cu and Pt. However, in the spectra for the de-alloyed PtCu films, the peak corresponding to $Cu^{2p}$ levels is substantially diminished indicating Pt surface enrichment. The kinetic-limited region from the oxygen-reduction curves obtained on Pt and de-alloyed PtCu electrodes in an RDE setup are shown in Figure 7. The true surface area of Pt and the dealloyed PtCu electrode were estimated from the hydrogen desorption charge in the cyclic voltammograms obtained in $N_2$-saturated 0.1 M $HClO_4$. Based on these measurements, the real surface area of the dealloyed PtCu electrode was estimated to be 16% larger than that of the Pt electrode; the current densities in Figure 7 are reported with respect to the real surface area. The de-alloyed PtCu electrode exhibits an enhanced performance towards ORR compared to that of the Pt electrode. The reduction in overpotential was measured at ten current densities between 1 and 2 $mA/cm^2_{rsa}$; the average shift in the potential was $24.42 \pm 0.01$ mV. This performance enhancement is similar to what was reported in the literature [3, 8, 9, 11, 43-47].

Stress-thickness measurements are carried out on 100 nm PtCu thin-film electrodes while they are de-alloyed between -0.05 V and +1.15 V *vs*. SHE in an Ar-saturated 0.1 M $HClO_4$ electrolytic solution. The CVs and the stress-thickness evolution are shown in Figure 8a and 8b, respectively. For the sake of clarity, only data corresponding to cycles 1, 10, 50, 100, 150, 200 and 250 are shown. The CVs are similar to those previously shown in Fig. 5(a). The anodic peak in current density around +0.7 V, which we attribute to Cu dissolution, steadily decreases with the number of cycles. It is interesting to note that some features are unique to the first cycle. No selective dissolution of Cu occurs and the anodic current observed at more positive potentials is larger than that normally attributed to Pt oxidation. This suggests that contaminants are oxidized from the surface prior to activating Cu dissolution [48]. The first anodic sweep shows a stress response that is also different from the other cycles. The shape of the stress curve is fairly similar to that of Pt that appears in the literature [34]; however, the magnitude of the stress change for the alloy is about 30% that of pure Pt. Over this potential range the stress response is primarily due to electrocapillarity; *i.e.*, charge-induced surface stress and the adsorption/desorption of species on the Pt surface [49]. On the return sweep, the stress shows a peak at about 0.7 V. This peak is accentuated as de-alloying continues and becomes a prominent feature on all subsequent cycles, appearing at 0.9 V on the anodic sweep and 0.7 V on the cathodic return sweep. Although we cannot ascribe a particular feature of the stress response to Cu dissolution or the re-arrangement of Pt, it is clear that this peak in the stress response is the direct result of de-alloying.

Another prominent feature of the stress-thickness profiles is the systematic shift towards compressive stress values as de-alloying progresses. This is indicated by the arrow in Figure 8b. The stress-thickness evolution in any CV cycle is a superposition of the stress-thickness induced due to de-alloying and that due to electrocapillary effects. As such, the stress-thickness due exclusively to de-alloying can be obtained by measuring the shift in stress-thickness from that of the first CV cycle at any potential in the scan range. Although there is some dependence of the shift with the potential value, we have chosen to measure the shift at the start of the CV cycle (-0.05 V *vs.* SHE). The stress-thickness evolution due to de-alloying, as shown in Figure 9, is compressive and increases rapidly during the first few cycles. The net change in stress-thickness reaches a plateau of about -0.7 to -0.95 N/m after 10 cycles.

The de-alloying process is reported to form a strained Pt-rich film in contact with the underlying PtCu alloy [6]. As such the associated stress response can be examined in a manner similar to that which has been developed for epitaxial thin film growth onto foreign metal substrates. The surface stress difference associated with geometrically strained overlayer growth, $\Delta F$, can be given by [50, 51]

$$\Delta F = \Delta f_o + \Delta f_{pzc} + \Delta f_i + \Delta\Sigma_{mf} h_f \qquad 2$$

where $\Delta f_o$ is the change in the intrinsic surface stress associated with the change in surface chemistry, $\Delta f_{pzc}$ is an electrocapillary term that reflects the change in the potential of zero charge (pzc) of the surface, $\Delta f_i$ is the interface stress between the overlayer and the substrate, and $\Delta\Sigma_{mf}h_f$ is the product of the misfit stress and the film thickness of the overlayer. If we assume that the PtCu alloy possesses surface properties that are intermediate to that of Cu and Pt, then one can expect the intrinsic surface stress to become more positive ($\Delta f_o \approx +1.0$ N/m) [52] as de-alloying causes the surface to become Pt-rich. Similar tensile stress should come from the electrocapillarity term ($\Delta f_{pzc}$). Compressive stress is generated as the potential of an electrode is set positive of its pzc [53, 54]. As the pzc is shifted positive toward that of Pt [55] the charge-induced compressive stress will be diminished, resulting in a net tensile stress. One can then expect the first two terms of Eqn. 2 to result in tensile stress. The interface stress between the PtCu alloy and the Pt-enriched layer is not known. This term can be either tensile or compressive but values for other fcc metal pairs are generally small, less than about 0.5 N/m [52, 56]. The strain in the enriched Pt layer reported by Strasser *et al.* for de-alloyed PtCu is about -3%, resulting in a large compressive misfit term in Eqn. 2. It is clear that the misfit stress dominates and is responsible for the net compressive stress that is observed experimentally during de-alloying. It should also be noted that if the Pt-rich layer forms in small islands rather than as a continuous film, then an additional size-dependent compressive strain can be added to the misfit strain [59].

As noted before, to obtain the stress in the de-alloyed Pt-layer, it is necessary to determine its thickness, which is done by measuring the change in the mass of the electrode through an electrochemical quartz crystal nano-balance (EQCN) study, following the method described in [26-29, 35]. The change in the mass of the electrode as measured in the EQCN experiments during two representative successive CV cycles is shown in Figure 10a. There are multiple processes that occur on the electrode surface that contribute to transient mass changes during a CV cycle, which are noted in figure 10a. The mass change denoted by A corresponds to Cu dissolution from PtCu; B corresponds to mass change due to reduction of Pt oxide and perchlorate desorption from PtCu and Pt surfaces; C corresponds to the net change in mass per CV cycle (due to Cu dissolution); and D corresponds to the mass of re-deposited Cu. The mass of dissolved Cu per unit area per cycle for the first ten CV cycles is shown in Figure 10b. Cu dissolution is rapid during the initial cycles, consistent with stress-thickness evolution in Figure 9.

As noted above, X-Ray diffraction (XRD) spectrum obtained on an as-prepared PtCu film indicates a predominantly (111) texture (Figure 2). The mass of Cu per monolayer per $cm^2$ of (111) PtCu alloy is 85.8 $ng/cm^2$/monolayer. From the EQCN measurements, at the end of 10 de-alloying cycles, the net mass of Cu removed is 210 $ng/cm^2$ (with an uncertainty of 22%), which is equivalent to de-alloying 2.4 ± 0.6 monolayers of PtCu. Assuming that the Pt atoms rearrange themselves to occupy the Cu sites and consolidate into a dense layer, the thickness of the de-alloyed Pt layer is equivalent to 1.2 ± 0.3 monolayers. Note that this estimate is similar to the measurements of Pt-skin thickness on de-alloyed PtCu nanoparticles reported by Bele *et al.* [18] and Hodnik *et al.* [19]. Taking the spacing between (111) planes in Pt to be 0.131 nm, the thickness of the de-alloyed Pt layer is 0.16 ± 0.04 nm. Taking the average stress-thickness value at the end of 10 de-alloying cycles to be 0.83 ± 0.13 N/m, the stress in the de-alloyed Pt is estimated to be 4.95 ± 1.3 GPa. Taking the biaxial modulus of Pt (111) to be 336 GPa (Young's modulus of 185 GPa, Poisson's ratio of 0.45), the biaxial compressive elastic strain in the de-alloyed layer is estimated to be 1.47 ± 0.4% [the cumulative error is calculated from the error in EQCM measurement (22%) and from the maximum variation in measured stress-thickness values between samples (13%)]. It should be noted that these estimates are only approximate since they involve extending bulk continuum concepts to very thin surface layers. To our knowledge, this is the first direct measurement of stress in the Pt-enriched layer due to surface de-alloying of PtCu.

*Computational results: Dissociative mechanism* – Figure 11 shows the free-energy diagram for ORR on Pt (111) *via* dissociative mechanism at $U = 1.23$ V, and its variation with biaxial strain applied to the Pt surface. Here $U$ is the electrode potential versus SHE. It can be seen that O* (see Figure 3) formation is exergonic (*i.e.,* downhill in free energy) but subsequent protonations of O* and of HO* are endergonic (uphill in free energy). For the free-energy landscape for oxygen reduction to be fully downhill the potential of the electrode would have to be lowered by a value equal to the larger of the two protonation barriers $\Delta G_L$ (D2 and D3 in Figure 11), also known as overpotential ($\eta$). The potential at which ORR will now take-off on the electrode is called the limiting potential ($U_L$) and is given by

$$\eta = 1.23 - U_L = \Delta G_L / e \qquad 3$$

where $e$ is the charge of an electron. Thus, $U_L$ can be considered as an ORR activity indicator (A) and has previously been shown to successfully predict the onset potentials in experiments [42, 58, 59]. As shown in Figure 11, compressive strain raises the free energy of the intermediates making them less stable and tensile strain lowers their free energy making them more stable. Note that strain has a stronger effect on the adsorption free energy of O* than that of HO*. It can be seen in Figure 12a that ΔG(D3) > ΔG(D2) for strains between -2.5% and 0.5% and, ΔG(D2) > ΔG(D3) for strains between 0.5% and 2.5%. Figure 12b shows the variation of $U_L$ with strain; it can be seen that $U_L$ increases with compressive strain (i.e., reduction in overpotential) and decreases with tensile strain (*i.e.*, increase in overpotential), which is consistent with the experimental observation on de-alloyed PtCu electrodes reported in Figure 7 as well as the observations in the literature [1-6].

*Associative mechanism* – The free energy diagram for ORR on biaxially strained Pt (111) according to associative mechanism at $U = 1.23$ V is shown in Figure 13. All four intermediates are destabilized by compression and stabilized by tension. The effect of the application of strain is larger on adsorption energies of $O_2$* and O* than that of HOO* and HO*. It can be seen that $O_2$* and O* formations are downhill in energy while the protonation of $O_2$*, O*, and HO* are uphill in energy. Over the entire strain range considered herein, the protonation of $O_2$* to HOO* (step A2 in the scheme shown in Figure 3) always has the largest free energy change. When a compressive biaxial strain is applied to Pt (111), ΔG (A2) decreases and it increases under tensile strain as shown in Figure 14(a). ORR activity calculated from equation (2) using ΔG (A2) leads to a variation with strain as illustrated in Figure 14(b). In other words, ORR activity *via* associative mechanism increases with compression, similar to that in the dissociative mechanism.

*Shift in potential with strain* – The variation of the change in the ORR, $U_L$, relative to that of unstrained Pt (111) with biaxial strain *via* associative and dissociative mechanisms is shown in Figure 15. Under compression, the rate of change in $U_L$ is -32.8 mV and -36.9 mV per 1% strain for the dissociative and associative mechanisms, respectively. Under tension, these numbers are -41.4 mV and -36.7 mV respectively. Fig. 15 also shows the experimental measurement (Fig. 7) of the shift in CV potential ($\Delta V$ = 24.42 ± 0.01 mV) for de-alloyed Pt/Pt-Cu films in which the Pt film was inferred to be under a compressive strain of 1.47%. Note that the computational results are in good agreement with the experimental measurement.

## 5. Conclusions

We report an experimental investigation in which stress evolution during de-alloying of PtCu catalyst films is measured directly by monitoring the change in substrate curvature in real time. In order to estimate the thickness of the de-alloyed layer, the mass of de-alloyed Cu was measured using an electrochemical quartz crystal nano-balance and it was shown to be 210 ± 46 $ng/cm^2$, which corresponds to a de-alloyed layer thickness of 1.2 ± 0.3 monolayers or 0.16 ± 0.04 nm. Real-time substrate-curvature measurements during electrochemical de-alloying show development of compressive stress in the de-alloyed Pt layer and its magnitude was estimated to be 4.95 ± 1.3 GPa, which corresponds to a biaxial elastic strain of 1.47 ± 0.4%. Further, the strained Pt layer shows enhanced catalytic activity towards ORR compared to that of unstrained polycrystalline Pt as measured by a reduction in overpotential of 24.42 ± 0.01 mV, which agrees with similar measurements in the literature. We have also carried out DFT calculations on biaxially strained Pt surface to estimate the influence of strain on the ORR overpotential by considering both dissociative and associative mechanisms. The theoretically predicted change in overpotential with strain was similar in magnitude for both mechanisms and was consistent with the direction and magnitude of the experimentally measured CV shift for the Pt/PtCu dealloyed electrode. The predictive capability of the computations, supported by its agreement with the experimental measurement is expected to provide a framework for the design of core-shell catalysts.

## 6. Acknowledgements

The authors at Brown University gratefully acknowledge financial support from the Army Research Office - Multidisciplinary University Research Initiative (ARO-MURI Grant # W911NF-11-1-0353) on "Stress Controlled Catalysis *via* Engineered Nanostructures." Cantilever metallization by electron beam evaporation was performed at the Center for Nanoscale Science and Technology NanoFab at the National Institute of Standards and Technology. Certain trade names are mentioned for experimental information only, and in no case does it imply a recommendation or endorsement by NIST.

**TABLES**

**Table 1: Parameters used for the stress and energy analyses presented in this study.**

| Parameter | Definition | Value | Comments |
|---|---|---|---|
| $D_{psd}$ | Distance between PSD and electrode | 81.2 cm | Measured |
| $E_s$ | Young's modulus of cantilever substrate | 72.9 GPa | Ref. [60] |
| $h_s$ | Thickness of cantilever substrate | 100 μm | Measured |
| $L$ | Submerged length of cantilever | 20 mm | Measured |
| $n_{H_2O}$ | Refractive index of electrolyte | 1.33 | - |
| $n_{air}$ | Refractive index of air | 1 | - |
| $\upsilon_s$ | Poisson's ratio of cantilever substrate | 0.28 | Ref. [60] |

## FIGURE CAPTIONS

Figure 1: Schematic experimental setup for real-time measurement of stress in the de-alloyed Pt layer on a Pt-Cu film. (a) Laser path is shown along with the components constituting the cantilever-deflection measurement system. The electrochemical cell containing 0.1 M HClO4 and the cell components are shown on the right. (b) Schematic showing the de-alloyed Pt/PtCu thin-film electrode on Ti/Au-coated borosilicate glass substrate, which serves as the cantilever..

Figure 2: XRD spectrum of sputter-deposited PtCu (1:1) thin film on Ti/Au-coated quartz crystal substrate, revealing a predominantly (111) texture.

Figure 3: Elementary steps involved in dissociative and associative mechanisms for the reduction of molecular oxygen to water on a catalyst surface.

Figure 4: Structural geometries of (a) Pt (111) surface, and (b) O*, (c) HO*, (d) O2*, and (e) HOO* adsorbed Pt (111) surfaces, respectively.

Figure 5: (a) Schematic showing the Pt/PtCu structure prepared via voltammetric de-alloying of sputter-deposited PtCu (1:1) alloy. Note that the layers are not drawn to scale. (b) The first four de-alloying CV cycles are shown. These CVs are obtained in 0.1M HClO4 (de-aerated with UHP Ar) at a scan rate of 20 mV/s and at 298 K and 1 atm. (c) The 600th CV cycle is shown. The arrows in both figures indicate the direction in which the potential is swept.

Figure 6: XPS spectra of PtCu thin film (in the energy range of Cu 2p and Pt 4f signals) before and after 600 cycles of voltammetric de-alloying.

Figure 7: Current density with respect to the real surface area corresponding to the cathodic (negative direction) sweep on Pt and Pt/PtCu thin-film (rotating-disk) electrodes in 0.1 M HClO4 (saturated with UHP O2) at a scan rate of 20 mV/s are shown. A positive shift of 24.42 ± 0.01 mV in the overpotential toward oxygen-reduction reaction is observed for the dealloyed PtCu electrode.

Figure 8: Selected (a) cyclic voltammogram and (b) stress-thickness evolution during de-alloying of a PtCu thin-film electrode in 0.1 HClO4 electrolyte saturated with UHP Ar are shown. Scan rate: 250 mV/s.

Figure 9: Evolution of stress-thickness as a function of number of dealloying cycles for two PtCu cantilever electrodes that also show the experimental variability.

Figure 10: (a) Mass change (□m) corresponding to the third and fourth de-alloying CV cycles on PtCu thin-film electrode in 0.1 M HClO4 (de-aerated with UHP Ar) is shown. Phenomena resulting in electrode mass changes are labeled A through D and are described in the text. The CV sweep direction is indicated by the arrows. (b) Net loss in electrode mass for the first 10 CV cycles is shown.

Figure 11: Free energy diagram at $U$ = 1.23 V for oxygen reduction reaction *via* dissociative mechanism on Pt (111) biaxially strained by -2.5%, 0.0%, and +2.5%.

Figure 12: (a) Reaction free energies of protonation steps D2 and D3 in the scheme shown in Figure 2, and (b) activity in terms of the limiting potential ($UL$) at $U$ = 1.23 V for ORR *via* dissociative mechanism over ± 2.5% in-plane biaxially strained Pt (111) surfaces.

Figure 13: Free energy diagram at $U$ = 1.23 V for oxygen reduction reaction *via* associative mechanism on Pt (111) biaxially strained by -2.5%, 0.0%, and +2.5%.

Figure 14: (a) Reaction free energy of protonation step A2 in the scheme shown in Figure 2 and (b) activity in terms of the limiting potential ($UL$) at $U$ = 1.23 V for ORR *via* associative mechanism over ± 2.5% in-plane biaxially strained Pt(111) surfaces.

Figure 15: Shift in limiting potential from calculations for ORR on Pt (111) *via* dissociative and associative mechanisms as a function of in-plane biaxial strain (ex, ey) is shown along with experimentally determined potential-shift vs biaxial-strain for Pt/PtCu (filled-circle).

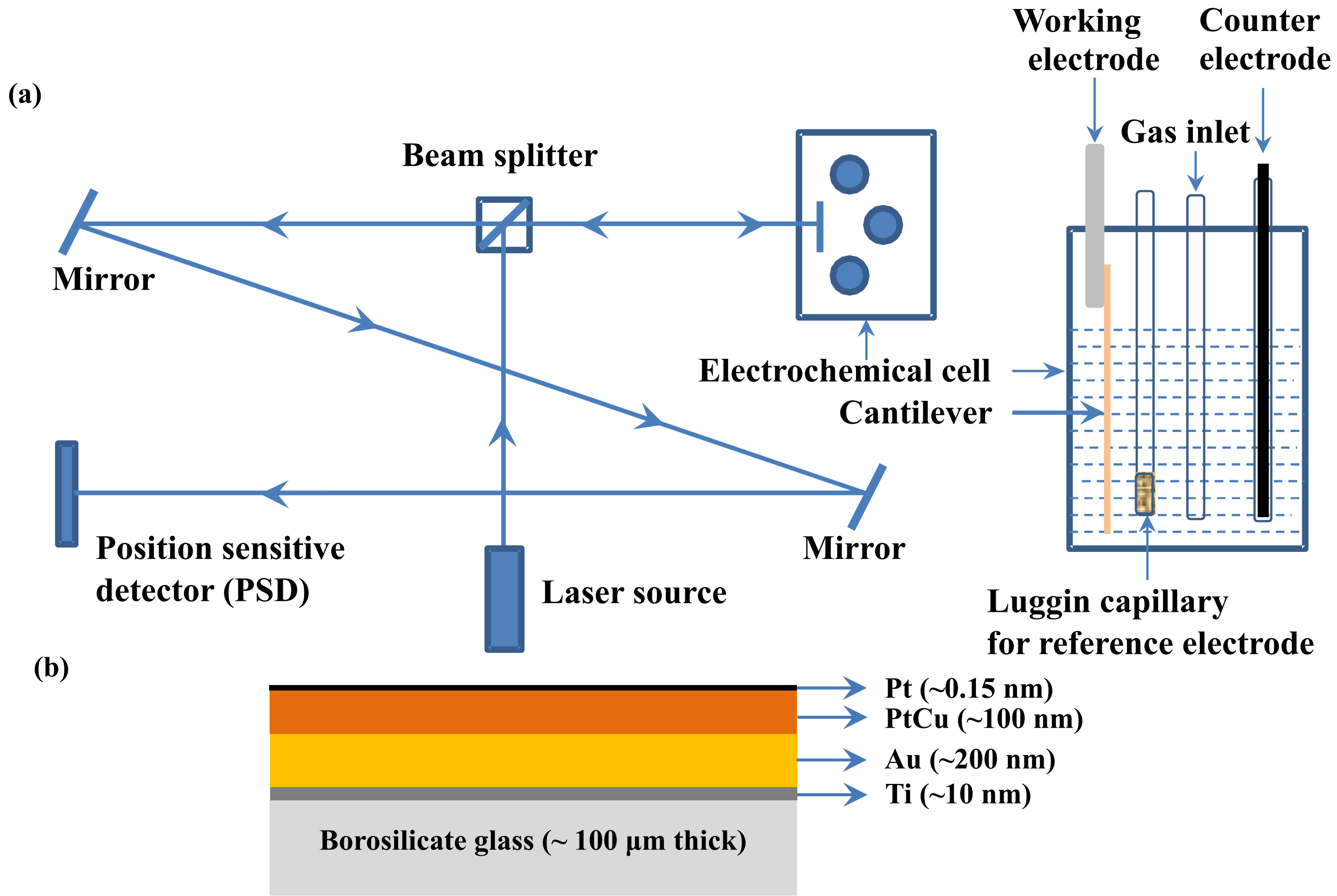

Figure 1: Schematic experimental setup for real-time measurement of stress in the de-alloyed Pt layer on a Pt-Cu film. (a) Laser path is shown along with the components constituting the cantilever-deflection measurement system. The electrochemical cell containing 0.1 M $HClO_4$ and the cell components are shown on the right. (b) Schematic showing the de-alloyed Pt/PtCu thin-film electrode on Ti/Au-coated borosilicate glass substrate, which serves as the cantilever.

**FIGURE 2**

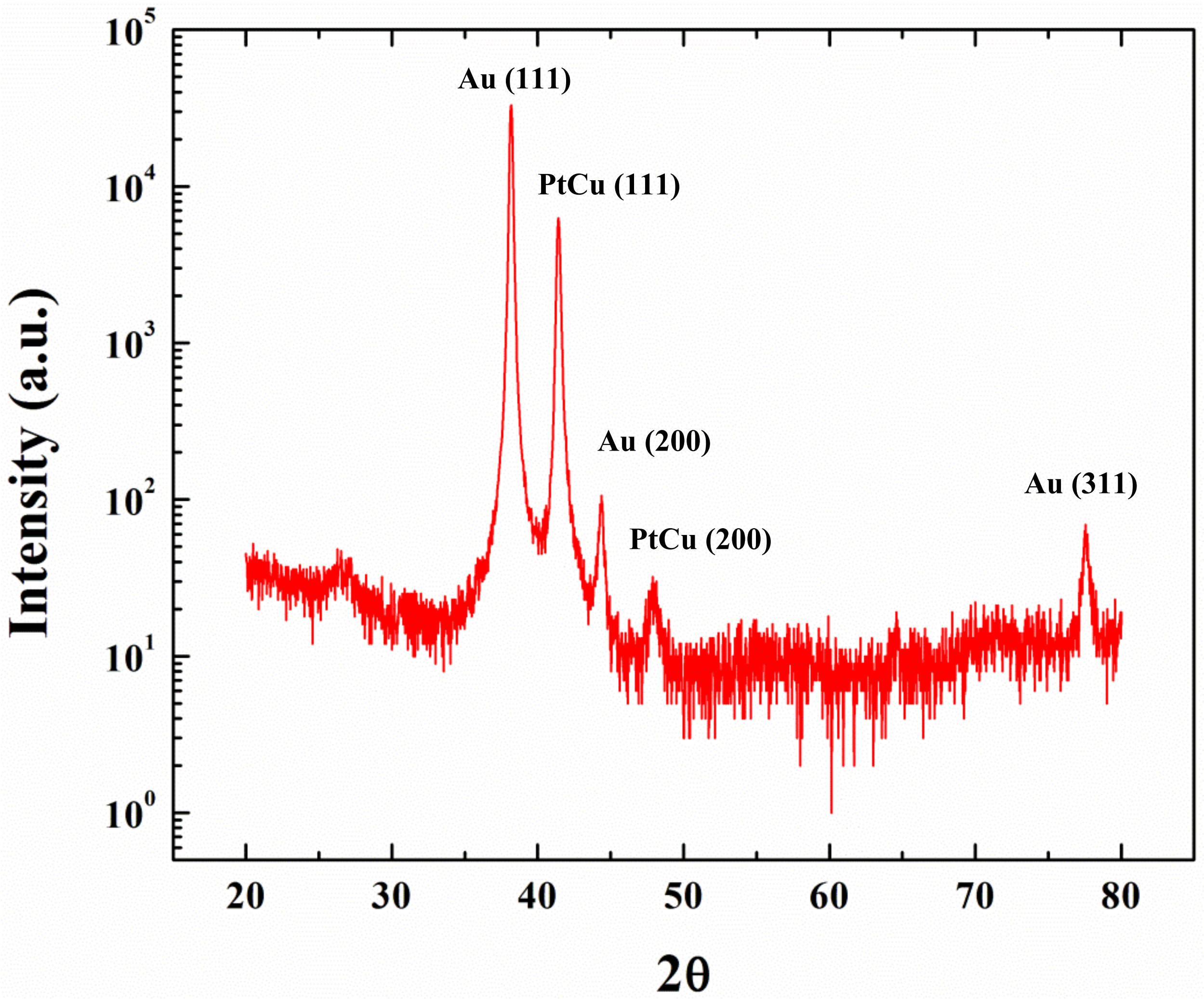

Figure 2: XRD spectrum of sputter-deposited PtCu (1:1) thin film on Ti/Au-coated quartz crystal substrate, revealing a predominantly (111) texture.

**FIGURE 3**

**Dissociative mechanism**

$$\tfrac{1}{2} O_2 (g) + * \rightarrow O* \quad (D1)$$

$$O* + H^+ + e^- \rightarrow HO* \quad (D2)$$

$$HO* + H^+ + e^- \rightarrow H_2O (l) + * \quad (D3)$$

**Associative mechanism**

$$O_2 (g) + * \rightarrow O_2* \quad (A1)$$

$$O_2* + H^+ + e^- \rightarrow HOO* \quad (A2)$$

$$HOO* + H^+ + e^- \rightarrow H_2O (l) + O* \quad (A3)$$

$$O* + H^+ + e^- \rightarrow HO* \quad (A4)$$

$$HO* + H^+ + e^- \rightarrow H_2O (l) + * \quad (A5)$$

Figure 3: Elementary steps involved in dissociative and associative mechanisms for the reduction of molecular oxygen to water on a catalyst surface. Note that * represents an adsorption site.

**FIGURE 4**

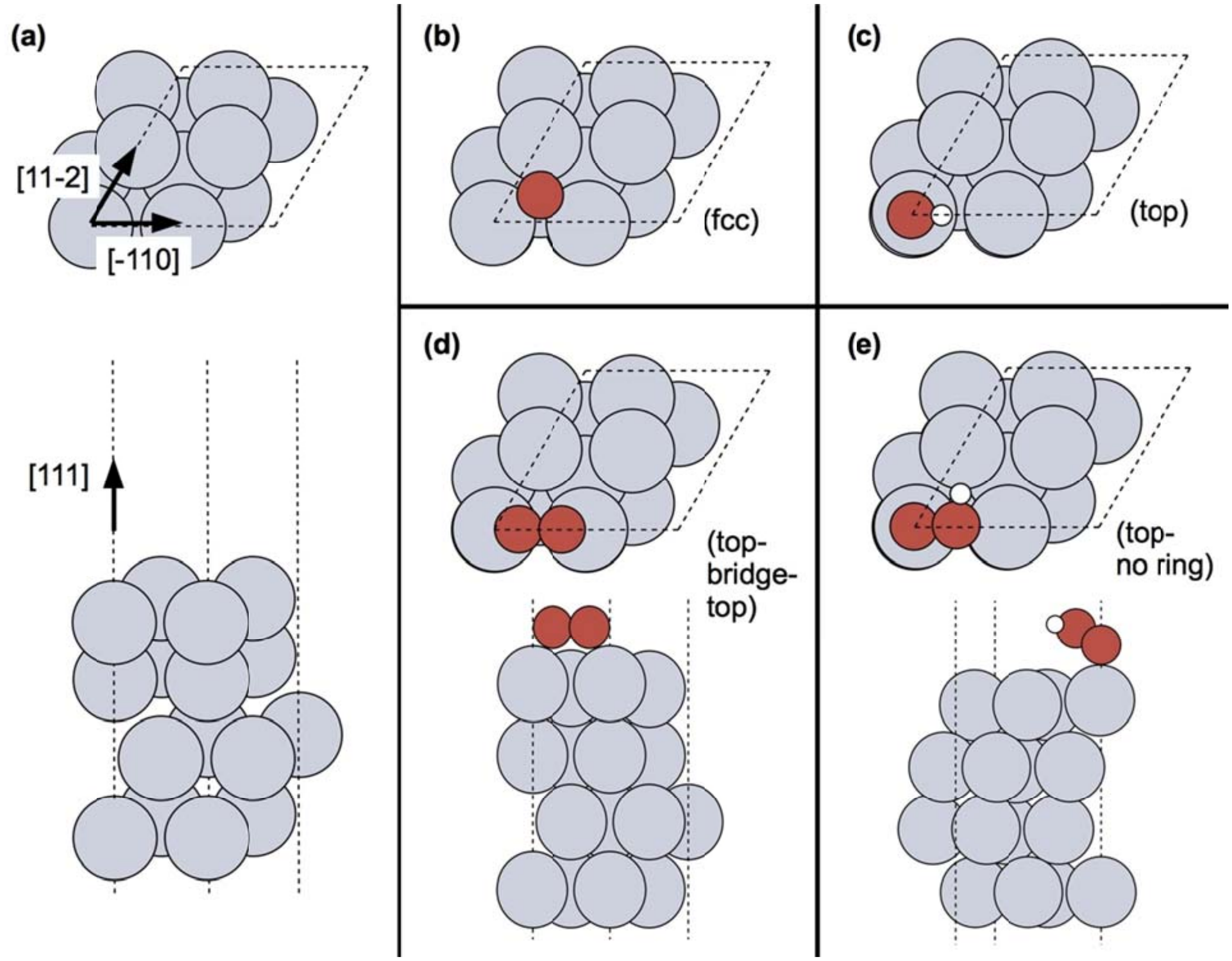

Figure 4: Structural geometries of (a) Pt (111) surface, and (b) O*, (c) HO*, (d) $O_2$*, and (e) HOO* adsorbed Pt (111) surfaces, respectively. The atoms in the descending order of size are Pt, O and H, respectively.

**FIGURE 5**

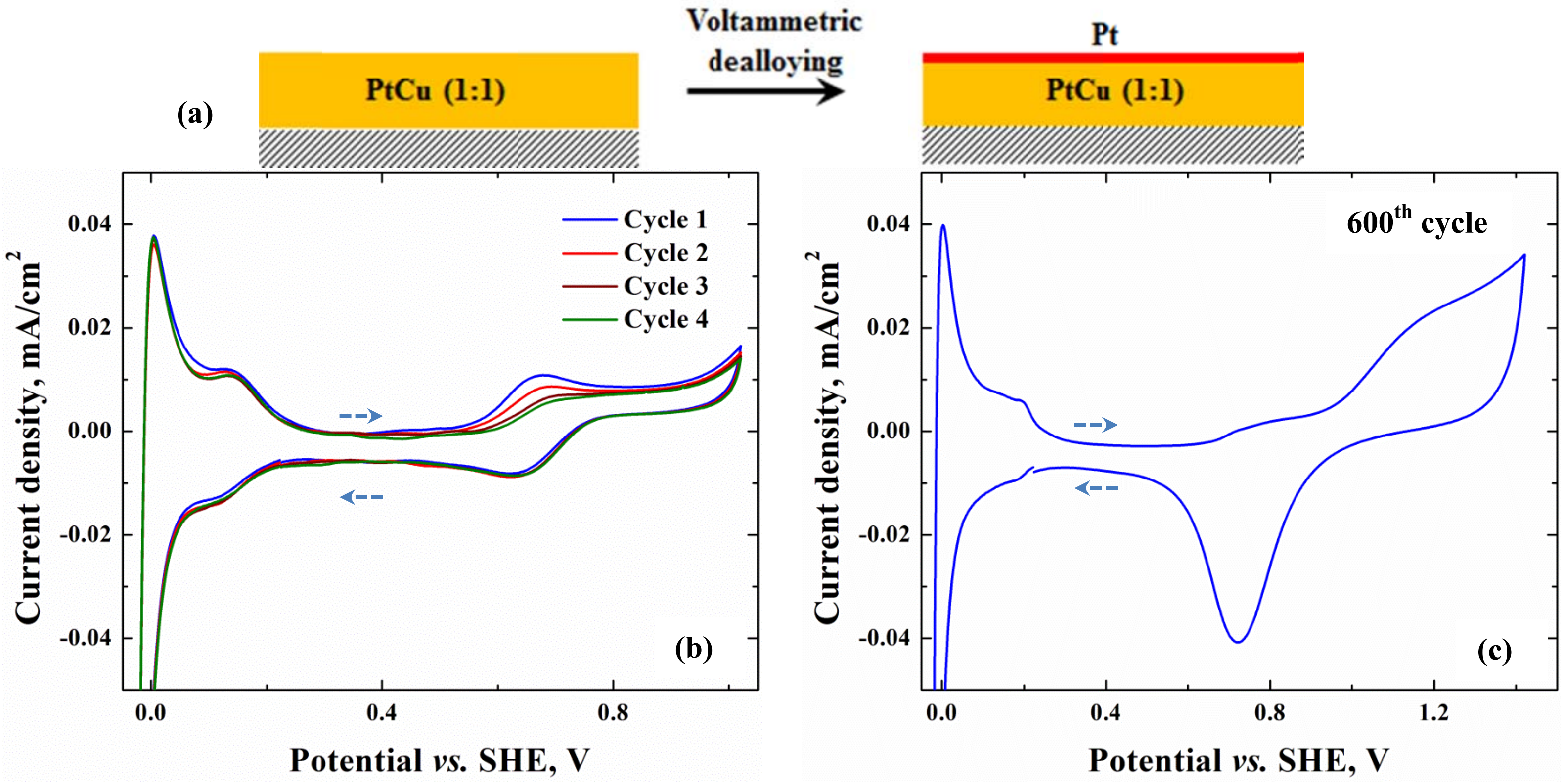

Figure 5: (a) Schematic showing the Pt/PtCu structure prepared *via* voltammetric de-alloying of sputter-deposited PtCu (1:1) alloy. Note that the layers are not drawn to scale. (b) The first four de-alloying CV cycles are shown. These CVs are obtained in 0.1M $HClO_4$ (de-aerated with UHP Ar) at a scan rate of 20 mV/s and at 298 K and 1 atm. (c) The $600^{th}$ CV cycle is shown. The arrows in both figures indicate the direction in which the potential is swept.

**FIGURE 6**

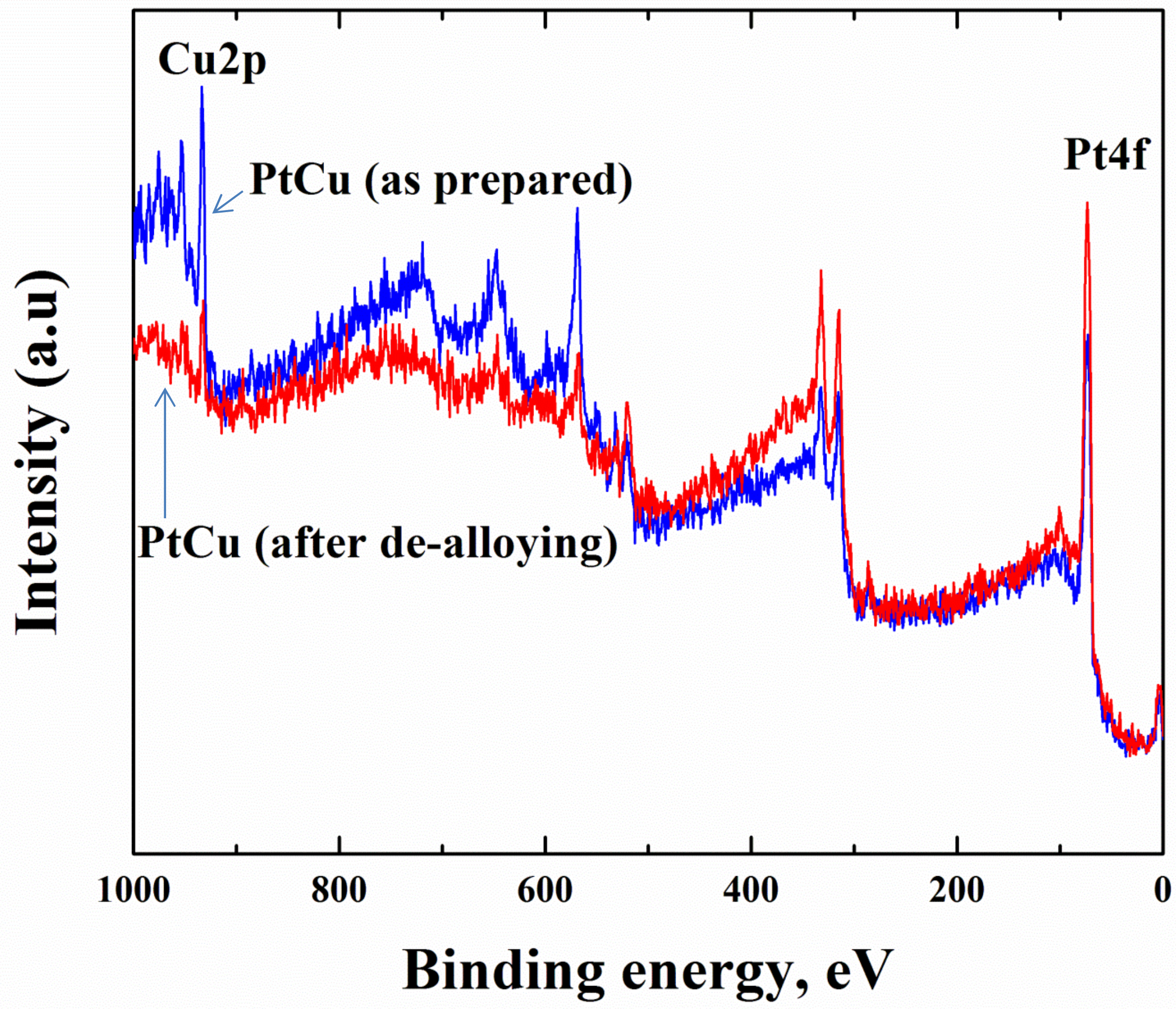

Figure 6: XPS spectra of PtCu thin film (in the energy range of Cu 2p and Pt 4f signals) before and after 600 cycles of voltammetric de-alloying.

**FIGURE 7**

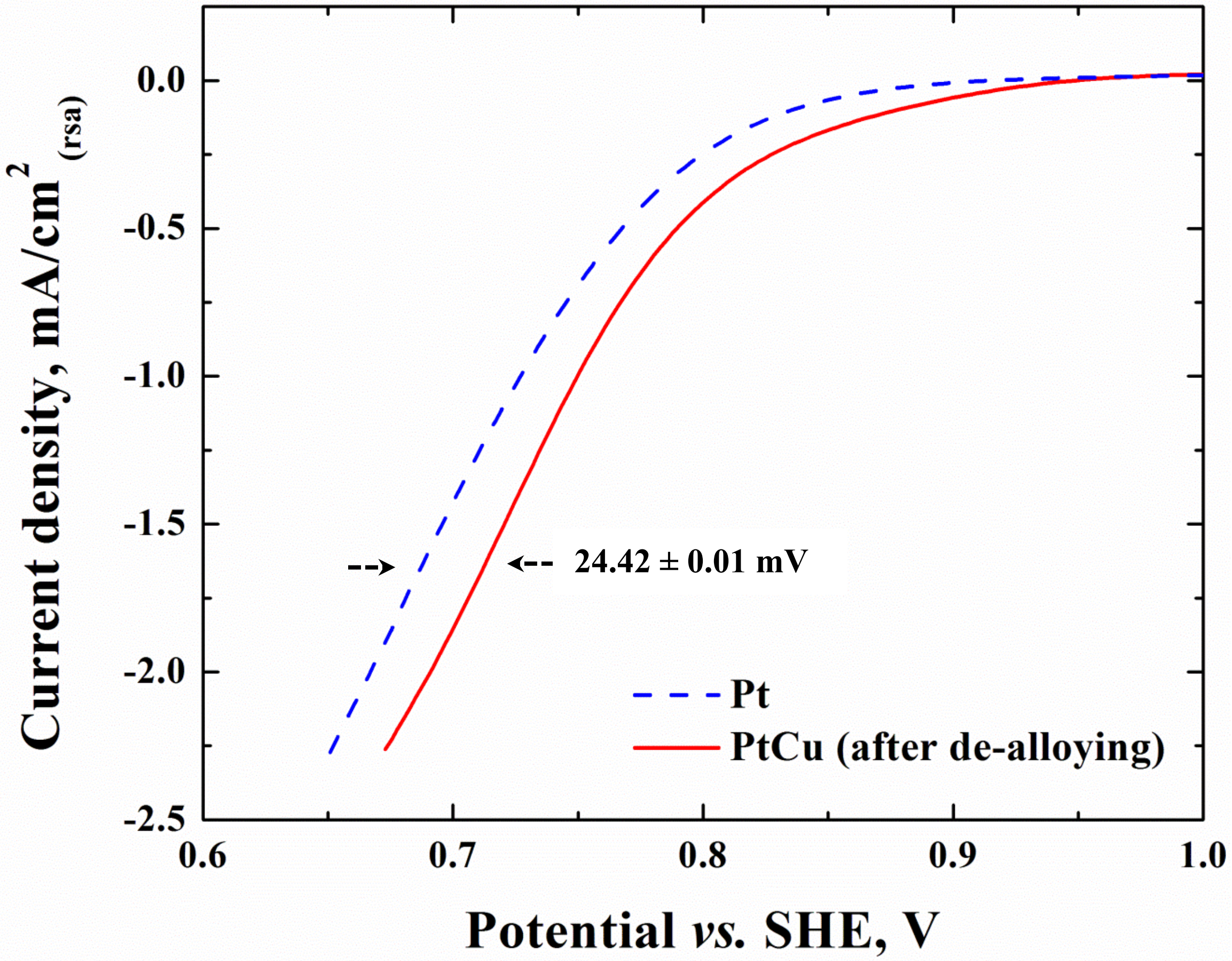

Figure 7: Current density with respect to the real surface area corresponding to the cathodic (negative direction) sweep on Pt and Pt/PtCu thin-film (rotating-disk) electrodes in 0.1 M $HClO_4$ (saturated with UHP $O_2$) at a scan rate of 20 mV/s are shown. A positive shift of 24.42 ± 0.01 mV in the overpotential toward oxygen-reduction reaction is observed for the dealloyed PtCu electrode. Note that the potential shift is an average value of potential shifts measured at multiple current densities between 1 and 2 $mA/cm^2_{rsa}$.

## FIGURE 8

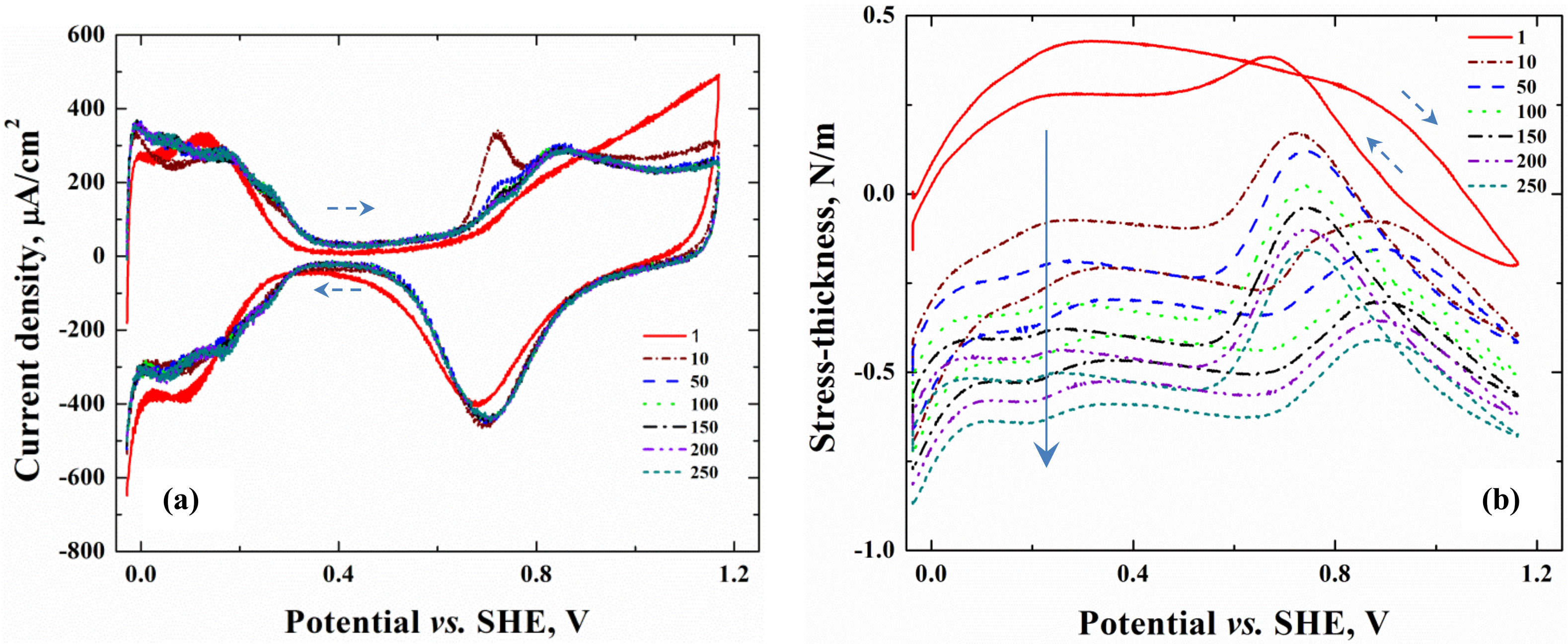

Figure 8: Selected (a) cyclic voltammogram and (b) stress-thickness evolution during de-alloying of a PtCu thin-film electrode in 0.1 $HClO_4$ electrolyte saturated with UHP Ar are shown. Scan rate: 250 mV/s.

**FIGURE 9**

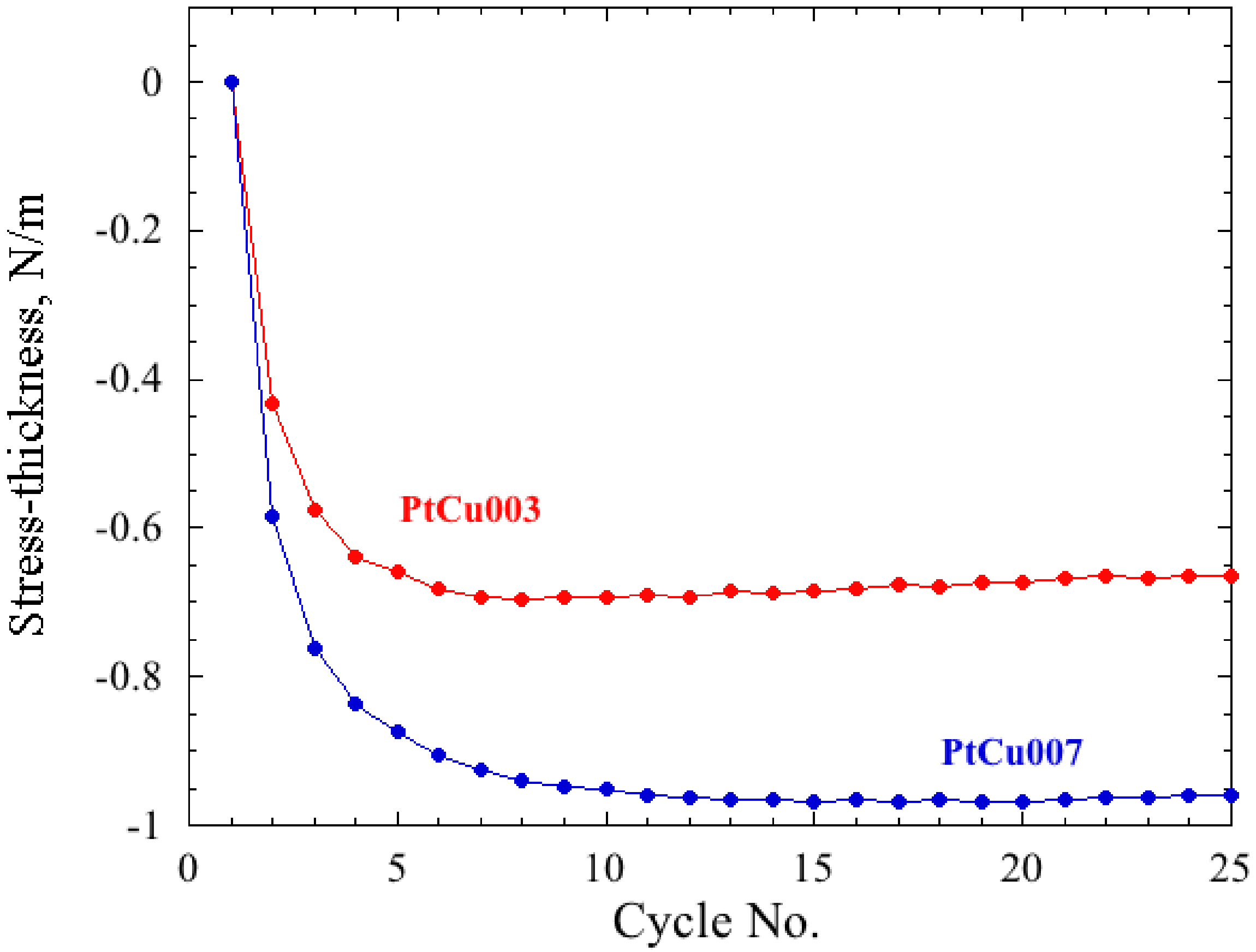

Figure 9: Evolution of stress-thickness as a function of number of dealloying cycles for two PtCu cantilever electrodes that also show the experimental variability.

**FIGURE 10**

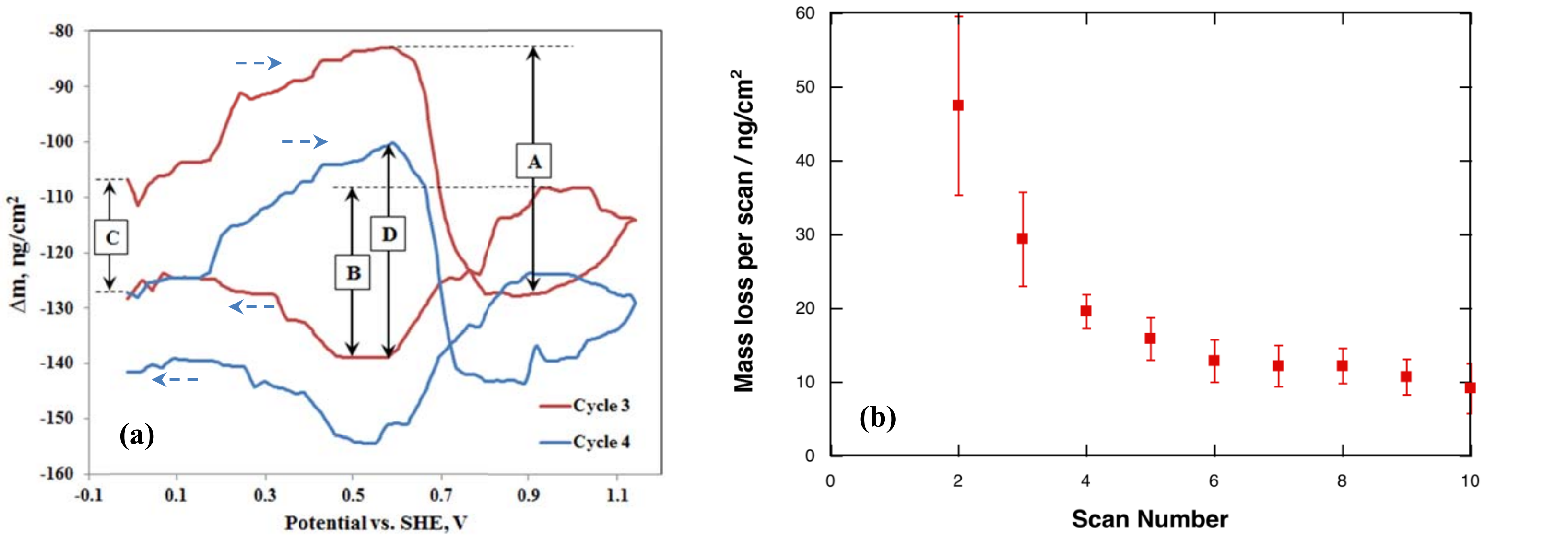

Figure 10: (a) Mass change (Δm) corresponding to the third and fourth de-alloying CV cycles on PtCu thin-film electrode in 0.1 M $HClO_4$ (de-aerated with UHP Ar) is shown. Phenomena resulting in electrode mass changes are labeled A through D and are described in the text. The CV sweep direction is indicated by the arrows. (b) Net loss in electrode mass for the first 10 CV cycles is shown.

**FIGURE 11**

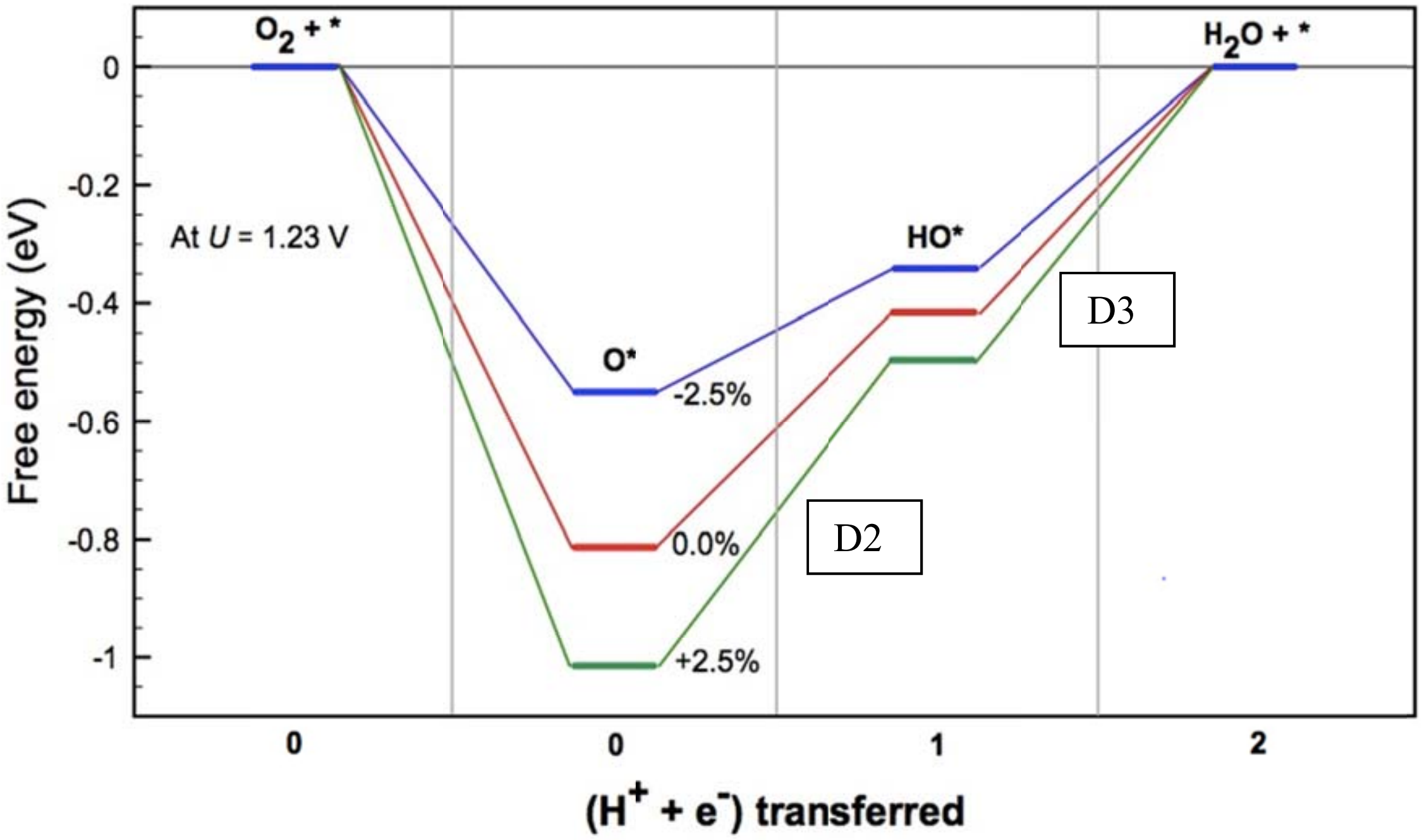

Figure 11: Free energy diagram at $U$ = 1.23 V for oxygen reduction reaction *via* dissociative mechanism on Pt (111) biaxially strained by -2.5%, 0.0%, and +2.5%.

**FIGURE 12**

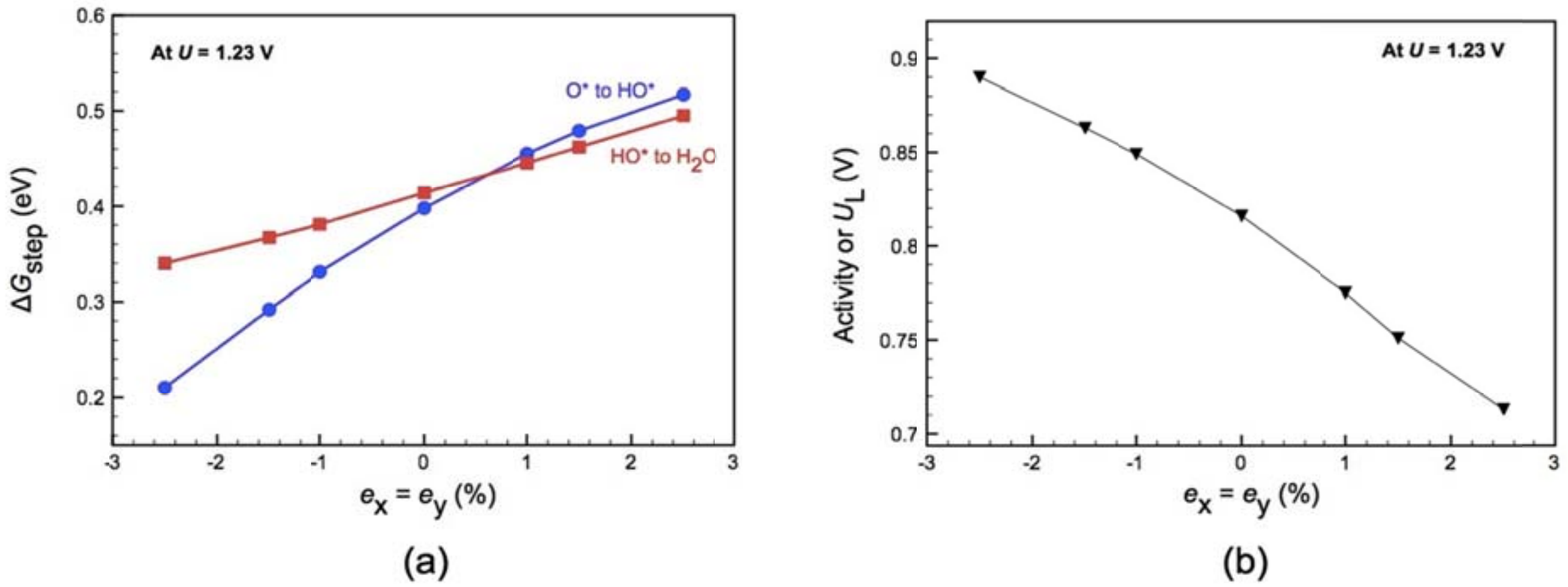

Figure 12: (a) Reaction free energies of protonation steps D2 and D3 in the scheme shown in Figure 2, and (b) activity in terms of the limiting potential ($U_L$) at $U$ = 1.23 V for ORR *via* dissociative mechanism over ± 2.5% in-plane biaxially strained Pt (111) surfaces.

**FIGURE 13**

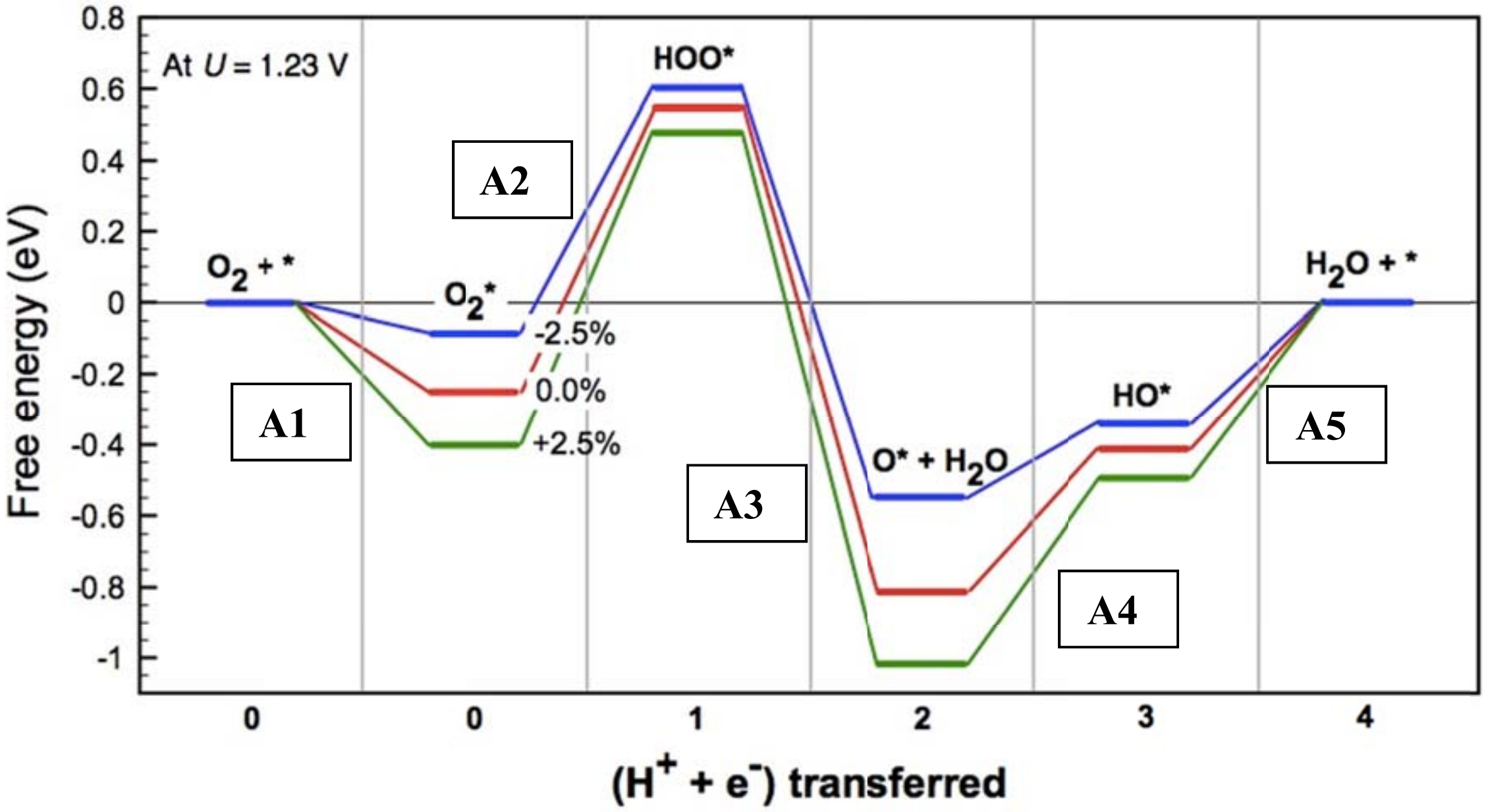

Figure 13: Free energy diagram at $U$ = 1.23 V for oxygen reduction reaction *via* associative mechanism on Pt (111) biaxially strained by -2.5%, 0.0%, and +2.5%.

**FIGURE 14**

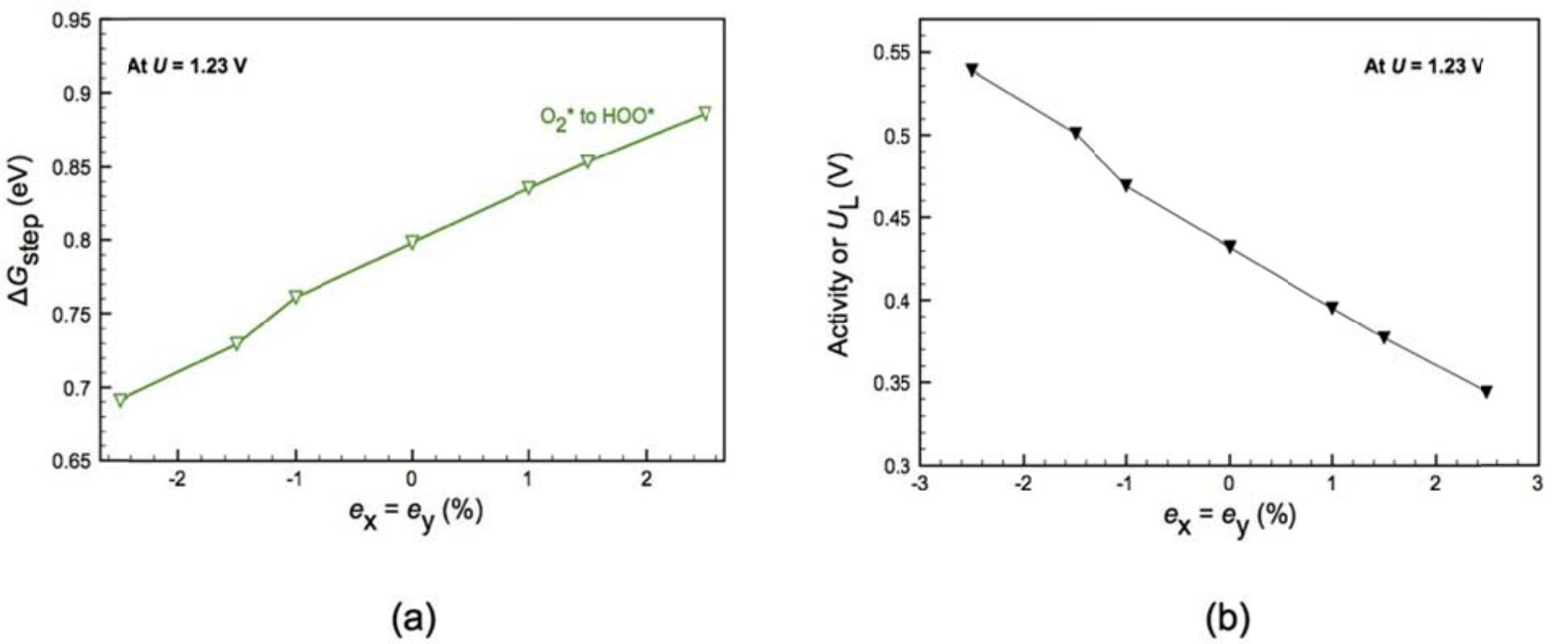

Figure 14: (a) Reaction free energy of protonation step A2 in the scheme shown in Figure 2 and (b) activity in terms of the limiting potential ($U_L$) at $U$ = 1.23 V for ORR *via* associative mechanism over ± 2.5% in-plane biaxially strained Pt(111) surfaces.

**FIGURE 15**

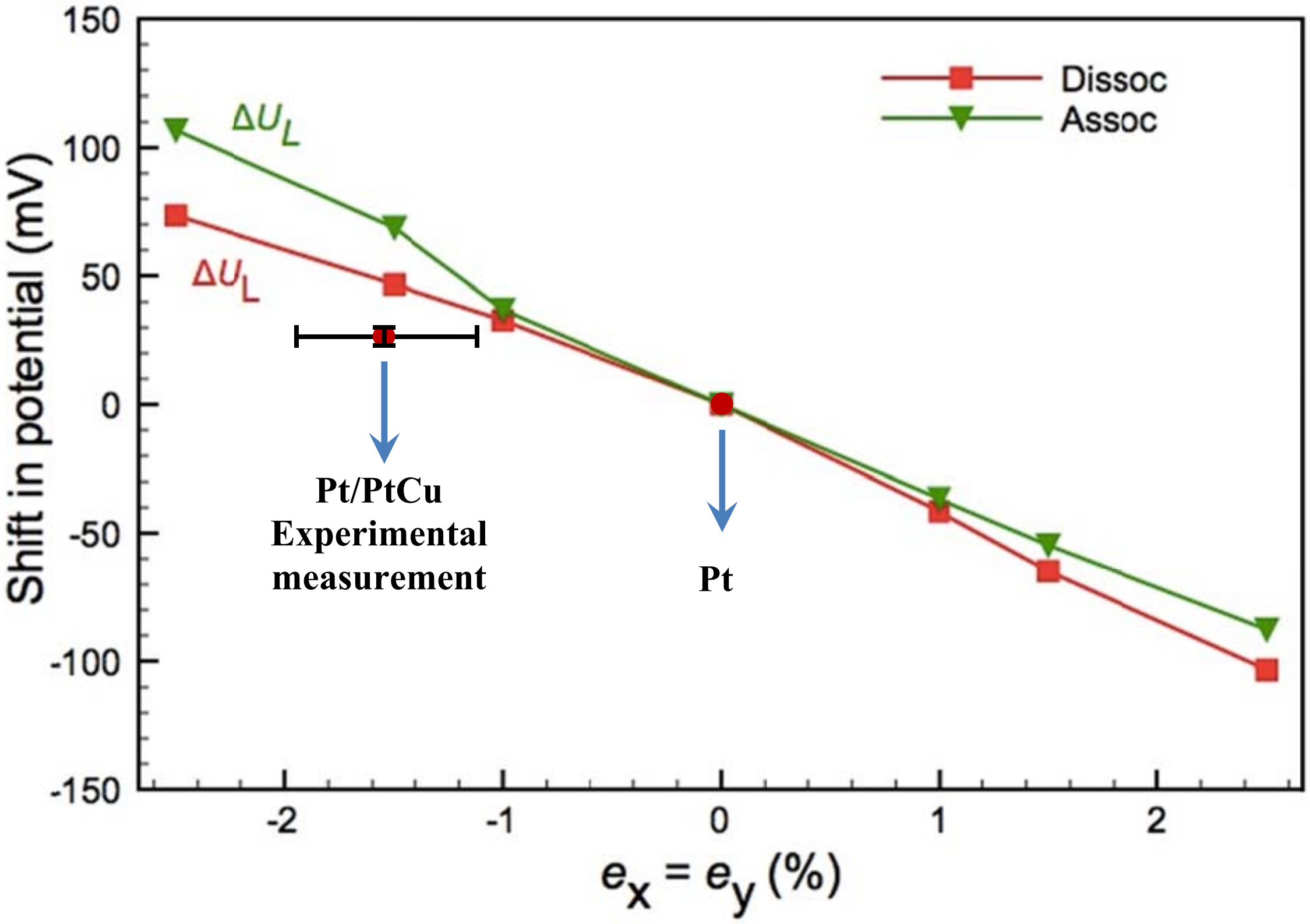

Figure 15: Shift in limiting potential from calculations for ORR on Pt (111) *via* dissociative and associative mechanisms as a function of in-plane biaxial strain ($e_x$, $e_y$) is shown along with experimentally determined potential-shift *vs* biaxial-strain for Pt/PtCu (filled-circle). The horizontal error bar indicates the cumulative error due to error in EQCM measurement and from the maximum variation in measured stress-thickness values between samples; the vertical error bar indicates the scatter in the measured shift in potential.

**References:**

1. Liu, Z.; Koh, S.; Yu, C.; Strasser, P. Synthesis, Dealloying, and ORR Electrocatalysis of PDDA-stabilized Cu-rich Pt Alloy Nanoparticles. *J. Electrochem. Soc.* **2007**, 154 (11), B1192-B1199.
2. Koh, S.; Strasser, P. Electrocatalysis on Bimetallic Surfaces: Modifying Catalytic Reactivity for Oxygen Reduction by Voltammetric Surface Dealloying. *J. Am. Chem. Soc.* **2007**, 129 (42), 12624-12625.
3. Mani, P.; Srivastava, R.; Strasser, P. Dealloyed Pt-Cu Core-shell Nanoparticle Electrocatalysts for Use in PEM Fuel Cell Cathodes. *J. Phys. Chem. C* **2008**, 112 (7), 2770-2778.
4. Strasser, P.; Koh, S.; Greeley, J. Voltammetric Surface Dealloying of Pt Bimetallic Nanoparticles: An Experimental and DFT Computational Analysis. *Phys. Chem. Chem. Phys.* **2008**, 10 (25), 3670-3683.
5. Neyerlin, K.C.; Srivastava, R.; Yu, C.; Strasser, P. Electrochemical Activity and Stability of Dealloyed Pt-Cu and Pt-Cu-Co Electrocatalysts for the Oxygen Reduction Reaction (ORR). *J. Power Sources* **2009**, 186 (2), 261-267.
6. Strasser, P.; Koh, S.; Anniyev, T.; Greeley, J.; More, K.; Yu, C.; Liu, Z.; Kaya, S.; Nordlund, D.; Ogasawara, H.; Toney, M.F.; Nilsson, A. Lattice-strain Control of the Activity in Dealloyed Core–Shell Fuel Cell Catalysts. *Nat. Chem.* **2010**, 2 (6), 454-460.
7. Dutta, I.; Carpenter, M.K.; Balogh, M.P.; Ziegelbauer, J.M.; Moylan, T.E.; Atwan, M.H.; Irish, N.P. Electrochemical and Structural Study of a Chemically Dealloyed PtCu Oxygen Reduction Catalyst. *J. Phys. Chem. C* **2010**, 114 (39), 16309-16320.
8. Yang, R.Z.; Leisch, J.; Strasser, P.; Toney, M.F. Structure of Dealloyed $PtCu_3$ Thin Films and Catalytic Activity for Oxygen Reduction. *Chem. Mater.* **2010**, 22 (16), 4712-4720.
9. Oezaslan, M.; Strasser, P. Activity of Dealloyed $PtCo_3$ and $PtCu_3$ Nanoparticle Electrocatalyst for Oxygen Reduction Reaction in Polymer Electrolyte Membrane Fuel Cell. *J. Power Sources* **2011**, 196 (12), 5240-5249.
10. Yu, X.; Wang, D.; Peng, Q.; Li, Y. High Performance Electrocatalyst: Pt-Cu Hollow Nanocrystals. *Chem. Comm.* **2011**, 47 (28), 8094-8096.
11. Oezaslan, M.; Hasche, F.; Strasser, P. $PtCu_3$, PtCu and $Pt_3Cu$ Alloy Nanoparticle Electrocatalysts for Oxygen Reduction Reaction in Alkaline and Acidic Media. *J. Electrochem. Soc.* **2012**. 159 (4), B444-B454.
12. Jayasayee, K.; Rob Van Veen, J.A.; Manivasagam, T.G.; Celebi, S.; Hensen, E.J.M.; de Bruijn, F.A. Oxygen Reduction Reaction (ORR) Activity and Durability of Carbon Supported PtM (Co, Ni, Cu) Alloys: Influence of Particle Size and Non-noble Metals. *Appl. Cat. B: Environ.* **2012**, 111-112, 515-526.
13. Hodnik, N.; Bele, M.; Hocevar, S. New Pt-skin Electrocatalysts for Oxygen Reduction and Methanol Oxidation Reactions. *Electrochem. Comm.* **2012**, 23, 125-128.
14. Marcu, A.; Toth, G.; Srivastava, R.; Strasser, P. Preparation, Characterization and Degradation Mechanisms of PtCu Alloy Nanoparticles for Automotive Fuel Cells. *J. Power Sources* **2012**, 208, 288-295.
15. Yang, R.Z.; Bian, W.; Strasser, P.; Toney, M.F. Dealloyed $PdCu_3$ Thin Film Electrocatalysts for Oxygen Reduction Reaction. *J. Power Sources* **2013**, 222, 169-176.

16. Wang, M.; Zhang, W.; Wang, J.; Minett, A.; Lo, V.; Liu, H.; Chen, J. Mesoporous Hollow PtCu Nanoparticles for Electrocatalytic Oxygen Reduction Reaction. *J. Mater. Chem. A* **2013**, 1 (7), 2391-2394.
17. Jeyabharathi, C.; Hodnik, N.; Baldizzone, C.; Meier, J.C.; Heggen, M.; Phani, K.L.N.; Bele, M.; Zorko, M.; Hocevar, S.; Mayrhofer, K.J.J. Time Evolution of the Stability and Oxygen Reduction Reaction Activity of PtCu/C Nanoparticles. *ChemCatChem* **2013**, 5 (9), 2627-2635.
18. Bele, M.; Jovanovič, P.; Pavlišič, A.; Jozinović, B.; Zorko, M.; Rečnik, A.; Chernyshova, E.; Hočevar, S.; Hodnik, N.; Gaberšček, M. A Highly Active $PtCu_3$ Intermetallic Core-shell, Multilayered Pt-skin, Carbon Embedded Electrocatalyst Produced by a Scale-up Sol-gel Synthesis. *Chem. Comm.* **2014**, 50 (86), 13124-13126.
19. Hodnik, N.; Jeyabharathi, C.; Meier, J.C.; Kostka, A.; Phani, K.L.; Rečnik, A.; Bele, M.; Hočevar, S.; Gaberšček, M.; Mayrhofer, K.J.J. Effect of Ordering of $PtCu_3$ Nanoparticle Structure on the Activity and Stability for the Oxygen Reduction Reaction. *Phys. Chem. Chem. Phys.* **2014**, 16 (27), 13610-13615.
20. Kongstein, O.E.; Bertocci, U.; Stafford, G.R. *In situ* Stress Measurements During Copper Electrodeposition on (111)-textured Au. *J. Electrochem. Soc.* **2005**, 152 (3), C116-C123.
21. Rouya, E.; Stafford, G.R.; Beauchamp, C.; Floro, J.A.; Kelly, R.G.; Reed, M.L.; Zangari, G. *In situ* Stress Measurements During Electrodeposition of Au-Ni Alloys. *Electrochem. Solid State Lett.* **2010**, 13 (11), D87-D90.
22. Shin, J.W.; Bertocci, U.; Stafford, G.R. Stress Response to Surface Alloying and Dealloying During Underpotential Deposition of Pb on (111)-Textured Au. *J. Phys. Chem. C* **2010**, 114 (17), 7926-7932.
23. Shin, J.W.; Bertocci, U.; Stafford, G.R. *In situ* Stress Measurement During Hydrogen Sorption on Ultrathin (111)-Textured Pd Films in Alkaline Electrolyte. *J. Electrochem. Soc.* **2011**, 158 (7), F127-F134.
24. Shin, J.W.; Bertocci, U.; Stafford, G.R. *In situ* Stress Measurement During Electrodeposition of $Ni_xPt_{1-x}$ Alloys. *J. Electrochem. Soc.* **2012**, 159 (8), D479-D485.
25. Stafford, G.R.; Beauchamp, C.R. *In situ* Stress Measurements During Al UPD onto (111)-textured Au from $AlCl_3$-EMImCl Ionic Liquid. *J. Electrochem. Soc.* **2008**, 155 (5), D408-D413.
26. Stafford, G.R.; Bertocci, U. *In situ* Stress and Nanogravimetric Measurements During Hydrogen Adsorption/Absorption on Pd Overlayers Deposited onto (111)-Textured Au. *J. Phys. Chem. C* **2009**, 113 (30), 13249-13256.
27. Stafford, G.R.; Bertocci, U. *In situ* Stress and Nanogravimetric Measurements During Underpotential Deposition of Pd on (111)-Textured Au. *J. Phys. Chem. C* **2009**, 113 (1), 261-268.
28. Stafford, G.R.; Bertocci, U. *In situ* Stress and Nanogravimetric Measurements During Underpotential Deposition of Pb on (111)-Textured Au. *J. Phys. Chem. C* **2007**, 111 (47), 17580-17586.
29. Stafford, G.R.; Bertocci, U. *In situ* Stress and Nanogravimetric Measurements During Underpotential Deposition of Bismuth on (111)-Textured Au. *J. Phys. Chem. B* **2006**, 110 (31), 15493-15498.
30. Xu, D.; Sriram, V.; Ozolins, V,; Yang, J.-M.; Tu, K.N.; Stafford, G.R.; Beauchamp, C. *In situ* Measurements of Stress Evolution for Nanotwin Formation During Pulse Electrodeposition of Copper. *J. Appl. Phys.* **2009**, 105 (2), 023521.

31. Xu, D.; Sriram, V.; Ozolins, V,; Yang, J.-M.; Tu, K.N.; Stafford, G.R.; Beauchamp, C. Erratum: *In situ* Measurements of Stress Evolution for Nanotwin Formation During Pulse Electrodeposition of Copper. *J. Appl. Phys.* **2010**, 108 (9), 099901.
32. Zangmeister, C.D.; Bertocci, U.; Beauchamp, C.R.; Stafford, G.R. *In situ* Stress Measurements During the Electrochemical Adsorption/desorption of Self-assembled Monolayers. *Electrochim. Acta* **2008**, 53 (23), 6778-6786.
33. Lafouresse, M.C.; Bertocci, U.; Beauchamp, C.R.; Stafford, G.R. Simultaneous Electrochemical and Mechanical Impedance Spectroscopy Using Cantilever Curvature. *J. Electrochem. Soc.* **2012**, 159 (10), H816-H822.
34. Lafouresse, M.C.; Bertocci, U.; Stafford, G.R. Dynamic Stress Analysis Applied to (111)-Textured Pt in $HClO_4$ Electrolyte. *J. Electrochem. Soc.* **2013**, 160 (9), H636-H643.
35. Shin, J.W.; Bertocci, U.; Stafford, G.R. Underpotential Deposition of Tl on (111)-Textured Au: *In Situ* Stress and Nanogravimetric Measurements. *J. Phys. Chem. C* **2010**, 114 (41), 17621-17628.
36. DACAPO and ASE are open-source codes available from the Department of Physics, Technical University of Denmark, at https://wiki.fysik.dtu.dk.
37. Bahn, S.R.; Jacobsen, K.W. An Object-oriented Scripting Interface to a Legacy Electronic Structure Code. *Comp. Sci. Engg.* **2002**, 4 (3), 56-66.
38. Vanderbilt, D. Soft Self-Consistent Pseudopotentials in a Generalized Eigen Value Formalism. *Phys. Rev. B* **1990**, 41 (11), 7892-7895.
39. Hammer, B.; Hansen, L.B.; Nørskov, J.K. Improved Adsorption Energetics Within Density-functional Theory Using Revised Perdew-Burke-Ernzerhof Functionals. *Phys. Rev. B* **1999**, 59 (11), 7413-7421.
40. Nørskov, J.K.; Rossmeisl, J.; Logadottir, A,; Lindqvist, L.; Kitchin, J.R.; Bligaard, T; Jónsson, H. Origin of the Overpotential for Oxygen Reduction at a Fuel-cell Cathode. *J. Phys. Chem. B* **2004**, 108 (46), 17886-17892.
41. Zhdanov, V.P.; Kasemo, B. Kinetics of Electrochemical $O_2$ Reduction on Pt. *Electrochem. Comm.* **2006**, 8 (7), 1132-1136.
42. Karlberg, G. S.; Rossmeisl, J.; Nørskov, J.K. Estimations of Electric Field Effects on the Oxygen Reduction Reaction Based on the Density Functional Theory. *Phys. Chem. Chem. Phys.* **2007**, 9 (37), 5158-5161.
43. Mani, P.; Srivastava, R.; Strasser, P. Dealloyed Binary $PtM_3$ (M = Cu, Co, Ni) and Ternary $PtNi_3M$ (M = Cu, Co, Fe, Cr) Electrocatalysts for the Oxygen Reduction Reaction: Performance in Polymer Electrolyte Membrane Fuel Cells. *J. Power Sources* **2011**, 196 (2), 666-673.
44. Oezaslan, M.; Hasche, F.; Strasser, P. Oxygen Electroreduction on $PtCo_3$, PtCo and $Pt_3Co$ Alloy Nanoparticles for Alkaline and Acidic PEM Fuel Cells. *J. Electrochem. Soc.* **2012**, 159 (4), B394-B405.
45. Yang, R.Z.; Strasser, P.; Toney, M.F. Dealloying of $Cu_3Pt$ (111) Studied by Surface X-ray Scattering. *J. Phys. Chem. C* **2011**, 115 (18), 9074-9080.
46. Yang, R.Z.; Strasser, P.; Toney, M.F. Surface X-ray Scattering Studies of $Cu_3Pt$ (111) Model Electrocatalysts. *Abs. Am. Chem. Soc.* **2011**, 241.
47. Viswanath, R.N.; Kramer, D.; Weissmüller, J. Adsorbate Effects on the Surface Stress-charge Response of Platinum Electrodes. *Electrochim. Acta* **2008**, 53 (6), 2757-2767.

48. Conway, B.E.; Angerstein-Kozlowska, H.; Sharp, W.B.A.; Criddle, E.E. Ultrapurification of Water for Electrochemical and Surface Chemical Work by Catalytic Pyrodistillation. *Anal. Chem.* **1973**, 45 (8), 1331-1336.
49. Feibelman, P.J. First-principles Calculations of Stress Induced by Gas Adsorption on Pt(111). *Phys. Rev. B* **1997**, 56 (4), 2175-2182.
50. Friesen, C.; Dimitrov, N.; Cammarata, R.C.; Sieradzki, K. Surface Stress and Electrocapillarity of Solid Electrodes. *Langmuir* **2001**, 17 (3), 807-815.
51. Trimble, T.; Tang, L.; Vasiljevic, N.; Dimitrov, N.; Van Schilfgaarde, M.; Friesen, C.; Thompson, C.V.; Seel, S.C., Floro, J.A.; Sieradzki, K. Anion Adsorption Induced Reveral of Coherency Strain. *Phys. Rev. Lett.* **2005**, 95 (16), 166106.
52. Gumbsch, P.; Daw, M. Interface Stresses and Their Effects on the Elastic Moduli of Metallic Multilayers. *Phys. Rev. B* **1991**, 44 (8), 3934-3938.
53. Haiss, W. Surface Stress of Clean and Adsorbate-Covered Solids. *Rep. Prog. Phys.* **2001**, 64 (5), 591-648.
54. Ibach, H. The Role of Surface Stress in Reconstruction, Epitaxial Growth and Stabilization of Mesoscopic Structures. *Surf. Sci. Rep.* **1997**, 29 (5-6), 195-263.
55. Trasatti, S. Structure of the Metal/electrolyte Solution Interface: New Data for Theory. *Electrochim. Acta* **1991**, 36 (11-12), 1659-1667.
56. Shull, A.L.; Spaepen, F. Measurement of Stress During Vapor Deposition of Copper and Silver Thin Films and Multilayers. *J. Appl. Phys.* **1996**, 80 (11), 6243-6256.
57. Bae, S.-E.; Gokcen, D.; Liu, P.; Mohammadi, P.; Brankovic, S.R. Size Effects in Monolayer Catalysis - Model Study: Pt Submonolayers on Au(111). *Electrocatal.* **2012**, 3 (3-4), 203-210.
58. Shi, C.; Hansen, H.A.; Lausche, A.C.; Nørskov, J.K. Trends in Electrochemical $CO_2$ Reduction Activity for Open and Close-packed Metal Surfaces. *Phys. Chem. Chem. Phys.* **2014**, 16 (10), 4720-4727.
59. Viswanathan, V.; Hansen, H.A.; Rossmeisl, J.; Nørskov, J.K. Unifying the 2e(-) and 4e(-) Reduction of Oxygen on Metal Surfaces. *J. Phys. Chem. Lett.* **2012**, 3 (20), 2948-2951.
60. Datasheet for 0060-0003-0001-GF-CA, Precision Glass and Optics, Santa Ana, California.